\documentclass[twocolumn,showpacs,preprintnumbers,amsmath,amssymb]{revtex4}
\usepackage{graphicx}
\begin{document}
\title{ Bose-Einstein condensation and Superfluidity of magnetoexcitons in Graphene}
\author{  Oleg L. Berman,$^{1}$ Yurii E. Lozovik,$^{2}$ and Godfrey Gumbs$^{1}$}
\affiliation{\mbox{$^{1}$Department of Physics and Astronomy,
Hunter College of the City University of New York,} \\
695 Park Avenue, New York, NY 10021 \\
\mbox{$^{2}$Institute of Spectroscopy, Russian Academy of
Sciences, 142190 Troitsk, Moscow Region, Russia}}

\date{\today}

\begin{abstract}
We propose experiments to observe Bose-Einstein condensation (BEC) and
superfluidity of  quasi-two-dimensional (2D) spatially indirect magnetoexcitons
in bilayer graphene. The magnetic field $B$ is assumed strong.
The energy spectrum of collective excitations, the sound spectrum as well as
the effective magnetic mass of magnetoexcitons
are presented in the strong magnetic field regime. The
superfluid density $n_S$ and the temperature of the
Kosterlitz-Thouless phase  transition $T_c$ are shown to be increasing
functions of the excitonic density $n$ but decreasing functions of
$B$ and the interlayer separation $D$. Numerical results are presented from
these calculations.

\end{abstract}


\maketitle

Indirect excitons  in coupled quantum wells (CQWs) in
the presence or absence of a magnetic field $B$ have been the subject
of recent experimental investigations
\cite{Snoke,Butov,Timofeev,Eisenstein}.
These systems are of particular interest because of the possibility of
Bose-Einstein condensation (BEC) and the superfluidity of indirect excitons
formed from electron-hole pairs. These may result in persistent electrical
currents in each QW or coherent optical properties and
Josephson junction phenomena
\cite{Lozovik,Birman,Littlewood,Vignale_drag,Berman}.
In high magnetic fields, two-dimensional (2D) excitons
survive in a substantially wider temperature range, as the
exciton binding energies increase with magnetic field \cite{Lerner,Paquet,Kallin,Yoshioka,Ruvinsky,Ulloa,Moskalenko}.

In this Letter we propose a new physical realization of magnetoexcitonic BEC
and superfluidity in bilayer graphene with spatially separated electrons and holes in high magnetic field. Recent technological advances have allowed the production of graphene,
 which is a 2D honeycomb lattice of carbon atoms that form the basic planar structure in graphite \cite{Novoselov1,Zhang1} Graphene has been attracting a great deal
of experimental and theoretical attention because of unusual properties in its bandstructure \cite{Novoselov2,Zhang2,Nomura,Jain}.
It is a gapless semiconductor with massless electrons and holes which have
been described as Dirac-fermions \cite{DasSarma}.   Since there is no gap
between the conduction and valence bands in graphene without magnetic
 field, the screening effects result in the absence of excitons in
graphene in the absence of  magnetic field.  A strong magnetic field
produces a gap since the energy spectrum becomes discrete  formed by
Landau levels. The gap reduces screening  and leads to the formation of
magnetoexcitons.

We consider two parallel graphene layers separated by an insulating
slab of SiO$_2$. The electrons in one layer and holes in the
other  can be controlled as in the experiments with
CQWs\cite{Snoke,Butov,Timofeev,Eisenstein} by laser pumping (far
infrared in graphene). The spatial separation of electrons and holes
in different graphene layers can be achieved by applying an external
electric field. Furthermore, the spatially separated electrons and
holes can be created by varying the chemical potential by using a
bias voltage between two graphene layers or between two gates
located near the corresponding graphene sheets. Indirect magnetoexcitons
are bound states of spatially separated electrons and holes
in an external magnetic field. The ratio of the external voltage
$V_{ext}$ to the interlayer separation $D$ required to create spatially
separated electrons and holes in graphene layers with the 2D
density $n =10^{11}cm^{-2}$ is given by
$V_{ext}/D = 4\pi e n D/\epsilon_b = 4.021 \times 10^4$ V/cm.
Here, $-e$ is the electron charge and $\epsilon_b = 4.5$ is
the dielectric constant of SiO$_2$. Since the critical electric
field $E_{cr}$ of the dielectric breakdown
for SiO$_2$ is $E_{cr} \approx 10^{6} V/cm$, we conclude
that the external electric field
for the spatially separated electrons and holes is less than the critical electric
field for dielectric breakdown in SiO$_2$.

A conserved quantity for an isolated electron-hole pair for graphene
in magnetic field $B$ is the exciton magnetic momentum $\hat{\mathbf{P}}$ defined as

\begin{eqnarray}
\hat{\mathbf{P}} =  -i\hbar\nabla_{e} -i\hbar\nabla_{h} +
\frac{e}{c}(\mathbf{A}_{e} - \mathbf{A}_{h}) - \frac{e}{c}
[\mathbf{B}\times (\mathbf{r}_{e} -\mathbf{r}_{h})]\
\label{Momentum}
\end{eqnarray}
for the Dirac equation\cite{Iyengar} as for the Schr\"odinger
equation \cite{Gorkov,Lerner,Kallin}. Here, $\mathbf{r}_{e}$ and $\mathbf{r}_{h}$
are 2D coordinate vectors of an electron and hole, respectively,
$\mathbf{A}_{e}$ and $\mathbf{A}_{h}$ are the corresponding vector potential of
an electron and hole.  The cylindrical gauge for vector potential is used with
$\mathbf{A}_{e(h)} = 1/2  [\mathbf{B}\times \mathbf{r}_{e(h)}]$.

Neglecting transitions between Landau levels for high magnetic fields, we
employ first-order perturbation theory to the Coulomb attraction
$V(r)=-e^{2}/(\epsilon_{b}\sqrt{r^{2} + D^{2}})$ between an
electron and hole.   Here,
$\mathbf{r} = \mathbf{r}_{e} - \mathbf{r}_{h}$.
We calculate the magnetoexciton energy
using the expectation value for an electron in Landau level $1$ and a hole
in level $1$. We have

\begin{eqnarray}
& & E_{1,1} (P) = \left\langle 0,0,\mathbf{P}\left|\hat{V}(r)
\right|0,0,\mathbf{P}\right\rangle
\nonumber\\
&+& \left\langle 0,1,\mathbf{P}\left|\hat{V}(r)
\right|0,1,\mathbf{P}\right\rangle  +
\left\langle 0,1,\mathbf{P}\left|
\hat{V}(r)\right|0,1,\mathbf{P}\right\rangle
\nonumber\\
&+&\left\langle 1,0,\mathbf{P}\left|
\hat{V}(r)\right|1,0,\mathbf{P}\right\rangle ,
\label{en4}
\end{eqnarray}
where $\left|\tilde{n},m,\mathbf{P}\right\rangle $ is an eigenfunction
of the non-relativistic Hamiltonian of a non-interacting
electron-hole pair defined in  \cite{Lerner,Ruvinsky}.
For small magnetic momentum satisfying $P \ll \hbar/r_B$ and
$P \ll \hbar D/r_B^2$, we obtain the following relations\cite{Ruvinsky}

\begin{eqnarray}
\langle\langle \tilde{n}m \mathbf{P}|\hat{V}(r)| \tilde{n}m \mathbf{P}\rangle\rangle = \mathcal{E}_{\tilde{n}m}^{(b)} + \frac{P^{2}}{2M_{\tilde{n}m}(B,D)}\ .
\label{ruv}
\end{eqnarray}
Substituting these results into Eq.\ (\ref{en4}), we obtain
the binding energy $\mathcal{E}_{B}^{(b)}(D)$ and the effective magnetic
mass $m_B(D)$ of a magnetoexciton with spatially separated electron
and hole in bilayer graphene as $\mathcal{E}_{B}^{(b)}(D) = \mathcal{E}_{00}^{(b)}(B,D) + 2 \mathcal{E}_{01}^{(b)}(B,D) + \mathcal{E}_{10}^{(b)}(B,D)$, and $m_B^{-1}(D)
= M_{00}^{-1}(B,D) + M_{01}(B,D)^{-1} + M_{10}^{-1}(B,D)$,
where constants $\mathcal{E}_{00}^{(b)}(B,D)$, $\mathcal{E}_{01}^{(b)}(B,D)$, $\mathcal{E}_{10}^{(b)}(B,D)$, $M_{00}(B,D)$, $M_{01}(B,D)$ and $M_{10}(B,D)$ depending on magnetic field $B$ and the interlayer separation $D$ in detail are given in
\ \cite{Ruvinsky}:

\begin{eqnarray}
\mathcal{E}_{00}^{(b)}(B,D) &=& -\mathcal{E}_{0}\exp\left[\frac{D^2}{2r_B^2}\right]
 \mathrm{erfc}\left[\frac{D}{\sqrt{2}r_B}\right] , \nonumber \\
  M_{00}(B,D) &=&  M_{0}\left[\left(1+\frac{D^{2}}{r_B^2}\right)\exp\left[\frac{D^{2}}{2r_B^2}\right]
 \mathrm{erfc}\left[\frac{D}{\sqrt{2}r_B}\right] \right.
\nonumber\\
&-&\left. \sqrt{\frac{2}{\pi}}\frac{D}{r_B}\right]^{-1} \ .
\label{m2}
\end{eqnarray}
Here, $r_{B} = \sqrt{\hbar c/(eH)}$ is the magnetic length; $c$ is the speed of light,
$\mathcal{E}_{0} = e^2/\epsilon_b r_B\sqrt{\pi/2}$,
$M_{0}=2^{3/2}\hbar^{2}\epsilon_b /(\sqrt{\pi}e^{2}r_B)$ and
$\mathrm{erfc}(z)$ is the complementary error function \cite{Ruvinsky}.
The radius of a magnetoexciton in the lowest Landau level
is given by $r_{1,1}(B) = 4 r_B$.

For large interlayer separation $D\gg r_B$, the asymptotic values of
the binding energy $\mathcal{E}_{B}^{(b)}(D)$ and the effective magnetic
mass $m_B(D)$ are $\mathcal{E}_{B}^{(b)}(B,D) = -4e^2/(\epsilon_b D)$,
and $m_B(D) = \epsilon_b D^{3}B^2/(4 c^{2})$. When
$D \ll r_B$, these quantities denoted by $\mathcal{E}_{0}$ and
$M_{0}$ are presented above. We can see that the effective magnetic mass
of an indirect magnetoexciton is approximately four times smaller
than in CQWs at the same $D$, $\epsilon_{b}$ and $\mathbf{B}$ \cite{Ruvinsky}.
The magnetoexcitonic energy  is approximately four times larger in
bilayer graphene than in CQWs. Measuring energies relative to
the binding energy of a magnetoexciton,  the dispersion
relation  $\varepsilon _{k}(P)$  of a magnetoexciton is quadratic
at small magnetic momentum, i.e., $P \ll \hbar/r_B$ and  $P \ll \hbar D/r_B^2$.
We have $\varepsilon _k({\bf P}) = P^2/(2m_{B k})$, where $m_{B k}$ is the
effective magnetic mass, dependent  on $B$,
 $D$  and the magnetoexcitonic quantum numbers
$k = \{n_{+},n_{-}\}$ for an electron in Landau level $n_{+}$
and a hole in level $n_{-}$.
Indirect magnetoexcitons, either in the ground state or
an excited state, have electrical dipole moments.
We treat these excitons as interacting  {\it parallel} electric dipoles.
 This is valid when $D$ is larger than the mean separation $\langle  r \rangle $
between an electron and hole
parallel to the graphene layers. We take into
account that at high magnetic fields $\langle  r \rangle \approx
Pr_B^2/\hbar $ with $\langle {\bf r} \rangle $ perpendicular to ${\bf P}$.
 Typical values of magnetic momenta are given by $P \sim \hbar \sqrt{n}$,
where $n$ is the 2D density of magnetoexcitons for a parabolic
dispersion relation.  Consequently,
 $D \gg \langle r \rangle$ is valid when $D \gg \sqrt{n}
r_B^2 $. Since electrons on a  graphene lattice can be in two
valleys, there are four types of excitons in bilayer
graphene. Due to the fact that all these types of excitons have identical
envelope wave functions and energies\cite{Iyengar}, we consider below only
excitons in one valley. Also, we use $n_{0} = n/4$ as the density of excitons
in one layer,  with $n$ denoting the total density of excitons.
For large electron-hole separation $D\gg r_B$,
transitions between  Landau levels due to the Coulomb electron-hole
attraction can be neglected, if the following condition is valid,
i.e., $E_{b} = 4 e^{2}/(\epsilon_b D) \ll \hbar v_{F}/r_B$.
This corresponds to high magnetic field $B$, large interlayer
separation $D$ and large dielectric constant of the insulator layer
between the graphene layers. In this notation, $v_{F} = \sqrt{3}at/(2\hbar)$
is the Fermi velocity of electrons. Also, $a =2.566 \AA$  is a lattice
constant, $t \approx 2.71 eV$ is the overlap integral between
nearest carbon atoms \cite{Lukose}.

The distinction between excitons and bosons is due to exchange
effects \cite{Berman}.  These effects for excitons
with spatially separated electrons  and holes in a dilute system
satisfying $na^{2}(B,D) \ll 1$
are suppressed due to the negligible overlap of the wave functions of two
excitons as a result of the potential barrier, associated with the
dipole-dipole repulsion \cite{Berman}.
Two indirect excitons in a dilute system
interact via $U(R) = e^2D^2/(\epsilon_b R^3)$, where $R$ is the
distance between exciton dipoles along the graphene layers.
 In high magnetic fields, the small parameter mentioned above
has the form
$\exp[-2\hbar ^{-1}(m_{B k})^{1/2}eD a^{-1/2}(B,D)]$. So at $T = 0$,
the dilute gas of magnetoexcitons, which  is a boson system,
form a Bose condensate \cite{Griffin}. Therefore, the
system of indirect magnetoexcitons can be treated by the diagrammatic
technique for a boson system.
For the dilute 2D magnetoexciton system with $n a^2(B,D) \ll 1$,
the sum of ladder diagrams is adequate.
For the lowest Landau level, we denote
$\varepsilon _{11}({\bf P}) = \varepsilon ({\bf P})$.
Using the orthonormality of the four-component wave functions
of the relative coordinate for a non-interacting pair of
an electron in Landau level $n_{+}$ and a hole in level
$n_{-}$ ($\tilde{\Phi}_{n_{+},n_{-}} (\mathbf{P} =\mathbf{0},\mathbf{r})$)\cite{Iyengar}
we obtain an  approximate equation for the vertex $\Gamma $ in
strong magnetic fields. Due to the orthonormality of the four-component
wave functions $\tilde{\Phi}_{n_{+},n_{-}} ({\bf 0,r})$ the projection
of the equation for the vertex in the ladder approximation for a
dilute system onto the lowest Landau level results in the scalar
integral equation which does not reflect the spinor nature
of the four-component magnetoexcitonic wave functions in graphene.
In high magnetic field, one can ignore transitions between
Landau levels and consider only the lowest Landau level states
$n_{+}=n_{-}=1$.  Since typically, the value
of $r$ is $r_B$, and $P \ll \hbar/r_B$ in this approximation,
the equation for the vertex in the magnetic momentum representation  $P$
for the lowest Landau level has the same form  (compare with
\cite{Lozovik})  as for a 2D boson
system in the absence of magnetic field, but with  the magnetoexciton magnetic
mass $m_{B}$ (which depends on $B$ and $D$) instead of the exciton mass
($M = m_{e} + m_{B}$) and magnetic momentum instead of inertial momentum:

\begin{eqnarray}
&&\Gamma (\mathbf{p},\mathbf{p}';P) = U (\mathbf{p} - \mathbf{p}') \nonumber\\
&& + \int_{}^{} \frac{d^{2} q}{(2\pi \hbar)^2} \frac{U (\mathbf{p} -
\mathbf{q}) \Gamma (\mathbf{q},\mathbf{p}';P)}{\frac{\kappa ^2}{m_{B}}
+\Omega -
\frac{\mathbf{P}^{2}}{4m_{B}} - \frac{q^2}{m_{B}} + i\delta},
 \hspace{0.2cm} \delta\to 0^+ \nonumber\\
\mu &=& \frac{\kappa ^2}{2m_{B}} = n_0\Gamma _{0}
= n_{0} \Gamma (0,0;0)\ ,
\label{Gamma_Int}
\end{eqnarray}
where $P = \{\mathbf{P},\Omega \}$, and $\mu $ is the chemical
potential of the system. Equation\ (\ref{Gamma_Int}) is valid at
parameter values which satisfy the condition
for validity of the perturbation theory applied to the calculation
of the magnetoexcitonic binding energy. The specific feature
of a 2D Bose system is connected with a logarithmic
divergence in the  2D scattering amplitude at
zero energy \cite{Lozovik,Berman}. A simple analytical solution
of Eq.\ (\ref{Gamma_Int}) for the chemical potential can be obtained
if $\kappa m_B e^{2} D^2/(\hbar^{3}\epsilon) \ll 1$.  In strong magnetic
fields at $D \gg r_{B}$ the exciton magnetic mass  is defined as $m_B \approx
\hbar^{2}\epsilon D^{3}/(4e^{2}r_{B}^{4})$. So the inequality $\kappa m_B e^{2} D^2/(\hbar^{3}\epsilon) \ll 1$ is valid if $D \ll (r_{B}^{4}/n^{1/2})^{1/5}$.
Consequently, the chemical potential $\mu $ is obtained as

\begin{eqnarray}
\mu =  \frac{\kappa ^2 }{2m_{B}}
= \frac{8\pi \hbar^{2}n}{8m_B \log \left[ \hbar^{4}
\epsilon^{2}/\left(2\pi n m_B^2 e^4 D^4\right) \right]} .
\label{Mu}
\end{eqnarray}
 At small magnetic momentum, the solution of Eq.(\ref{Gamma_Int})
corresponds to the sound spectrum of collective excitations
$\varepsilon (P)  = c_s P$. Here, the sound velocity
$c_s = \sqrt{\Gamma n/(4m_B)}= \sqrt{\mu/m_B}$, where
$\mu $ is guven by Eq.\ (\ref{Mu}). Since magnetoexcitons have a
sound spectrum of collective excitations at small magnetic momentum
$P$ due to the dipole-dipole repulsion, the magnetoexcitonic superfluidity
is possible at low temperature $T$ in bilayer graphene. This is so
since the sound spectrum satisfies the Landau criterium of
superfluidity \cite{Griffin}.

It can be shown that when $D = 0$, the interaction between two magnetoexcitons
in the lowest Landau level can be neglected in strong magnetic field \cite{Lerner}.
The magnetoexcitons constructed by spatially separated electrons and holes
in bilayer graphene at large interlayer separations $D \gg r_B$ form
a weakly interacting 2D non-ideal Bose gas with a dipole-dipole repulsion.
Thus, the phase transition from the normal to superfluid phase
is the Kosterlitz-Thouless transition \cite{Kosterlitz}.
The temperature of this phase transition
$T_c $ to the superfluid state
in a   2D  magnetoexciton system is determined from
$T_c = \pi \hbar ^2 n_S (T_c)/(2 k_B m_B)$, where $n_S (T)$ is the
superfluid density of the magnetoexciton system,  as a function
of  $T$, $B$, $D$ and $k_B$ is Boltzmann's constant.
The function $n_S (T)$  can be obtained from
$n_S = n/4 - n_n $, with $n$ the total density and $n_N$ the normal
component density.  To calculate the superfluid component density,
we find the total quasiparticle current  in a
reference  frame in which the superfluid component is at rest.
We determine the normal component density by the usual procedure
\cite{Griffin}. Suppose that the magnetoexciton system  moves with
a velocity $\mathbf{u}$. At nonzero temperatures dissipating quasiparticles
will appear in this system. Since their density is small at low
temperatures, one can assume that the gas of quasiparticles is an ideal
Bose gas. To calculate the superfluid component density,
we find the total current of quasiparticles in a
frame of reference in which the superfluid component is at rest.
We denote $\langle Pk | ... | Pk \rangle $ as $\langle  \cdots \rangle $.
Using the Feynman theorem for isolated magnetoexcitons,
we obtain the velocity \cite{Gorkov}
$\mathbf{v} = \left\langle\hat{\mathbf{v}}\right\rangle =
\left\langle \partial \hat{H}/\partial \mathbf{P}   \right\rangle =
 \partial \varepsilon _k (P)/ \partial \mathbf{P} = \mathbf{P}/m_{B k}$.
We obtain the mean total current of 2D magnetoexcitons in the coordinate system,
moving with a velocity ${\bf u}$ as
$\left\langle \mathbf{J} \right\rangle = \left\langle \mathbf{P} \right\rangle/m_{B}$.
Expanding the integrand to first order by $\mathbf{P}\mathbf{u}/(k_{B}T)$,
we have

\begin{eqnarray}\label{J_Tot}
\langle \mathbf{J} \rangle &=& - \frac{\mathbf{u}}{2m_{B}}
\int\frac{d\mathbf{P}}{(2\pi \hbar)^{2}}P^{2}
\frac{\partial f\left[\varepsilon (P)\right]}
{\partial \varepsilon}
\nonumber \\
 &=& \frac{3 \zeta (3) }{2 \pi \hbar^{2}}\frac{k_{B}^{3}T^3}{m_{B}c_s^4} \mathbf{u}\ ,
\end{eqnarray}
where $f\left[\varepsilon (P)\right] = \left(\exp\left[
\varepsilon (P)/(k_{B}T)\right] - 1\right)^{-1}$ is the Bose-Einstein
distribution function and $\zeta (z)$ is the Riemann zeta function
($\zeta (3) \simeq 1.202$). Then we define the normal component density
is\cite{Griffin} $\langle  \mathbf{J} \rangle = n_n  \mathbf{u}$.
Applying Eq.\ (\ref{J_Tot}), we obtain the expression for the normal
density $n_N$.
As a result,  we have for the superfluid density: $n_S(T) = n/4 - n_N(T)$.
It follows that the expression for the superfluid density $n_S$
in strong magnetic field for the proposed magnetoexciton
system differs from the analogous
expression in the absence of  magnetic field in semiconductor CQWs (compare
with Refs.\ \cite{Berman}) by replacing the total exciton mass
$M = m_{e} +m_{B}$ with the magnetoexciton magnetic mass $m_{B}$.

\begin{figure}
\includegraphics[width=2.9in]{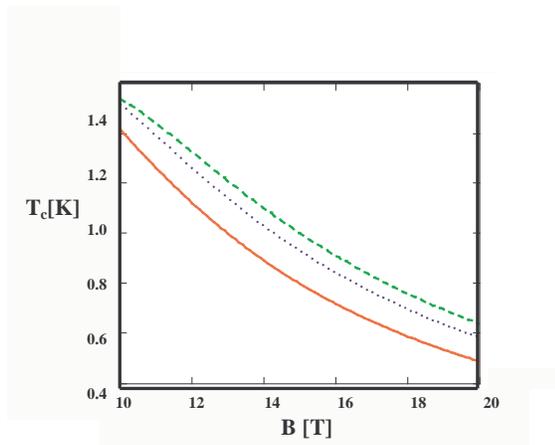}
\caption{(Color on line) Dependence of  Kosterlitz-Thouless transition temperature
$T_c = T_c (B)$  (in units $K$) versus magnetic field
for bilayer graphene separated by SiO$_2$.
with $\epsilon_b = 4.5$.
The magnetoexciton density $n = 4 \times 10^{11}cm^{-2}$.
Different interlayer separations $D$ are chosen: $D = 30 nm$ (solid
curve), $D = 28 nm$ (dotted curve), $D = 27 nm$ (dashed curve).}
\label{THD}
\end{figure}

In a 2D system, superfluidity of magnetoexcitons appears below the
Kosterlitz-Thouless transition temperature, where only coupled vortices are
present \cite{Kosterlitz}. Employing $n_S(T)$ for
the superfluid component, we obtain an
equation for the Kosterlitz-Thouless transition temperature
$T_c$ with solution

\begin{eqnarray}
 && T_c = \left[\left( 1 +
\sqrt{\frac{32}{27}\left(\frac{4m_Bk_BT_c^0}{\pi \hbar^{2} n}\right)^3 +
1} \right)^{1/3}  \right.
 \nonumber\\
 &&- \left. \left( \sqrt{\frac{32}{27}
\left(\frac{4 m_B k_BT_c^0}{\pi \hbar^{2} n}\right)^3 + 1} - 1 \right)^{1/3}\right]
\frac{T_c^{0}}{ 2^{1/3}}\ .
\label{tct}
\end{eqnarray}
Here, $T_c^{0}$ is an auxiliary quantity, equal to the
temperature at which the superfluid density vanishes in the
mean-field approximation, i.e., $n_S(T_c^{0}) = 0$,
$T_c^0 = k_B^{-1} \left(2 \pi \hbar^{2} n c_s^4 m_B/(12 \zeta (3))\right)^{1/3}$.
The temperature $T_c^0 = T_c^0(B,D)$ may be used to estimate
the crossover region where local superfluid density appearers
for magnetoexcitons  on a scale smaller or
of the order of the mean intervortex
separation in the system. The local superfluid density can manifest itself
in local optical or transport properties.
The dependence of $T_c$ on $B$ and $D$ is represented in Fig.\ \ref{THD}.
According to Eq.\ (\ref{tct}),  the temperature $T_c$
for the onset of superfluidity due to the Kosterlitz-Thouless
transition at {\it a fixed magnetoexciton density}
decreases as a function of magnetic field $B$ and interlayer separation $D$.
This is due to the increased $m_B$ as a functions  of $B$ and $D$.  The $T_c$
decreases as $B^{-{1}/{2}}$ at $D \ll r_B$ or as $B^{-2}$ when $D \gg r_B$.

In conclusion, we have studied BEC and superfluidity of magnetoexcitons
in two graphene layers with applied external voltage in  perpendicular
magnetic field. The superfluid density $n_S(T)$ and the temperature
of the Kosterlitz-Thouless phase transition to the superfluid state have
been calculated.  We have shown that at fixed exciton density $n$
the  Kosterlitz-Thouless temperature $T_c$ for the onset of superfluidity
of magnetoexcitons  decreases as a function of magnetic field like $B^{-1/2}$
at $D \lesssim r_B$ and as $B^{-2}$ when $D \gg r_B$.
We have shown that $T_c$ increases when the density $n$ increases
and decreases when the magnetic field $B$ and the interlayer separation
increase. The dependence of $T_c$ on $B$ and $D$ is presented in  Fig.\
 \ref{THD}. We also note that
the superfluidity of indirect magnetoexcitons in strong perpendicular magnetic
field in bilayer graphene is very interesting, because  the magnetoexcitons
in graphene are found to be more stable than in CQWs. Namely, the binding energy
of magnetoexcitons in graphene is four times greater than that
in CQWs with the same $D$, $\epsilon_{b}$ and $\mathbf{B}$. We consider
only the collective properties of excitons with electrons and holes from the same
valley. We note that there is no crossover  between Bose condensates
of  different types of excitons.

\acknowledgments
Yu.E.L. was supported by grants from RFBR and INTAS.
G.G. acknowledges partial support from the National Science
Foundation under grant \# CREST 0206162 as well as PSC-CUNY Award
\# 69114-00-38.

\end{document}